\documentclass{emulateapj}
\usepackage{hyperref,hyperref,amsmath,amssymb,times,graphicx,fleqn,color,mathtools}

\begin{document}

\newcommand{\bl}[1]{{\color{blue}#1}}
\newcommand{\ma}[1]{{\color{magenta}#1}}
\newcommand{\re}[1]{{\color{red}#1}}
\newcommand{\gr}[1]{{\color{green}#1}}
\newcommand{\ye}[1]{{\color{yellow}#1}}

\newcommand{\sciexp}[2]{{#1}\ensuremath{\,\times\,10^{#2}}}
\newcommand{\tsp}{\hspace{3mm}}
\newcommand{\mfp}{{m\!f\:\!\!p}}

\title{Kinetic simulations of electron pre-energization by magnetized collisionless shocks in expanding laboratory plasmas}

\author{K. V. Lezhnin\altaffilmark{1}}
\email{klezhnin@princeton.edu}
\author{W. Fox\altaffilmark{1,2}}
\author{D. B. Schaeffer\altaffilmark{1}}
\author{A. Spitkovsky\altaffilmark{1}}
\author{J. Matteucci\altaffilmark{1}}
\author{A. Bhattacharjee\altaffilmark{1,2}}
\author{K. Germaschewski\altaffilmark{3}}
\affil{$^1$Department of Astrophysical Sciences, Princeton University, Princeton, New Jersey 08544, USA}
%\affil{$^2$National Research Nuclear University MEPhI, 115409, Moscow, Russia}
\affil{$^2$Princeton Plasma Physics Laboratory, Princeton, New Jersey 08543, USA}
\affil{$^3$Department of Physics and Space Science Center, University of New Hampshire, Durham, New Hampshire 03824, USA}

%\date{\today}

\begin{abstract}
Collisionless shocks are common features in space and astrophysical systems where supersonic plasma flows interact, such as in the solar wind, the heliopause, and supernova remnants. Recent experimental capabilities and diagnostics allow detailed laboratory investigations of high-Mach-number shocks, which therefore can become a valuable way to understand shock dynamics in various astrophysical environments. {Using 2D particle-in-cell} simulations with a Coulomb binary collision operator, {we demonstrate the mechanism for generation of energetic electrons and experimental requirements for detecting this process in the laboratory high-Mach-number collisionless shocks}. We show through a parameter study that electron acceleration by magnetized collisionless shocks is feasible in laboratory experiments with {laser-driven expanding plasmas}. %Conditions for experimental observation of pre-accelerated electrons are formulated.

\bigskip

%\noindent Keywords: Magnetized collisionless shocks, Magnetic fields, Energized electrons, Particle-in-cell method
\end{abstract}

%\pacs{52.38.Kd, 52.65.Rr}
\maketitle

%\section{Introduction}

%Particle energization in astrophysical plasmas is one of the major problems of plasma astrophysics. 
Both Earth- and space-based detections of energetic particles spanning from MeV to EeV indicate that there are universal mechanisms for particle energization in astrophysical plasmas \citep{COSMICRAYS}. Two major plasma physics phenomena, magnetic reconnection \citep{REVIEW1,Bulanov2016} and collisionless shocks \citep{REVIEW2,REVIEW3}, are usually considered as main contributors to energetic particle populations. Magnetized collisionless shocks are naturally formed in many space environments with a pre-existing magnetic field, such as galaxy clusters, supernova remnants, and solar winds. The Fermi mechanism, commonly known as Diffusive Shock Acceleration \citep{Krymsky1977,DSA} (DSA), is a mechanism by which shocks can energize particles, creating a power-law energy spectrum of charged particles due to scattering of energized particles back and forth between upstream and downstream.
%by interaction with MHD waves, given sufficient pre-energization of particles such that their gyroradius is compared with typical wavelength of the Alfven waves. 

One of the major questions of electron energization by high-Mach-number magnetized collisionless shocks is the so-called ``injection problem'': in order to enter the Fermi energization cycle, particles must be pre-energized from the thermal level to have a gyroradius large enough to be able to scatter on upstream and downstream waves. Based on simulations (e.g., \citet{AmanoHoshino2007,AmanoHoshino2009,AmanoHoshino2010, Guo2014}), several different competing mechanisms have been proposed, but the need for a conclusive model still exists \citep{Katou2019}. Besides that, energetic particles are observed in the shock transition layer of moderate-level Alfv{\'e}n Mach number shocks with $M_{\rm A}\sim10$, even though turbulence in upstream and downstream may not be developed enough for lower shock speeds; thus, some other mechanism than DSA should be responsible for particle energization \citep{Matsumoto2015}. Moderate-level Alfv{\'e}n Mach number shocks are observed in the Earth magnetosphere, and the presence of %correlation between electrostatic waves and 
energized electrons was revealed from data by the Cassini satellite \citep{MastersNPHYS2013}.

Laboratory astrophysics experiments using expanding ablation plasmas from high power laser-solid interactions provide a platform for modeling of astrophysical processes, such as magnetic reconnection \citep{VULCAN,Shenguang2,Shenguang1, Rosenberg}, collisionless shocks \citep{Schaeffer2017,Schaeffer2019, UMEDA2019, Fiuza2020}, and Weibel instability \citep{Fox2013,Huntington2015}, allowing for detailed diagnostics \citep{Schaeffer2019} and controllable dimensionless parameters. Recently, magnetized collisionless shock formation with $M_{\rm A} \sim 15$, and upstream electron beta $\beta_e=8 \pi p_e/B^2 \sim 1$ was observed in the lab at the OMEGA laser facility \citep{Schaeffer2017}. Laboratory experiments with repeatable and controlled conditions and diagnostics which span local and global plasma scales can provide important information for solving the shock acceleration problem, benchmarking simulation, and ultimately providing important insight into interpreting energized particle observations in space and astrophysical plasmas. Simulations using the Plasma Simulation Code (PSC) \citep{Germaschewski2016}, which can match almost all dimensionless parameters of the system, allow detailed interpretation and predictions for the experiments. %allowing to match almost all dimensionless parameters of the system. 
A significant opportunity is therefore to design experiments to measure the efficiency of particle acceleration by shocks, and study how it relates to the geometry, plasma and field parameters, and microphysics of the shock.

In this Letter, we demonstrate with simulations the possibility of and requirements for observing electron pre-energization in magnetized shocks in the laboratory. The pre-acceleration is attributed to Shock Drift Acceleration (SDA), and we provide predictions for the first laboratory demonstration of this phenomenon. In contrast to typical shock simulations initiated by a moving simulation wall, we directly simulate self-consistent shock formation created by a laser-driven piston in plasma, including Coulomb particle collisions \citep{FoxPoP2018}, providing insights into the temporal behavior of particle acceleration in this strongly driven system. This leads to experimental requirements on the shock evolution time needed to distinguish particles accelerated at the shock from those generated by laser heating. Finally, we conduct simulations for a range of Mach numbers,
collisionalities %upstream plasma beta, 
and magnetic field inclinations, and find the optimal values for obtaining rapid particle acceleration at parameters which are not too distant from parameters obtained in recent laser driven shock experiments, and therefore may be possible in near future experiments.%, and obtain requirements on particle collisionality.

We perform simulations using the particle-in-cell code PSC, which has a module to simulate binary collisions and a heating operator to mimic laser-foil interaction \citep{Germaschewski2016,FoxPoP2018}. The 2-D simulation grid is in the $x-z$ plane, with $z$ being the shock propagation and primary ablation direction. In the simulations, a high density target is heated, which produces an energetic ablation plume (the ``piston'') expanding from a high density reservoir at density $n_{\rm ab}$ and temperature $T_{\rm ab}$, which interacts with and drives a shock in a low density magnetized background plasma (the ``upstream'') at density $n_{\rm up}$, $T_{\rm up}$, with magnetic field $B_0$ \citep{FoxPoP2018,SchaefferPoP2020}. In this work we simulate quasi-perpendicular shocks and therefore the initial magnetic field is oriented out-of-plane, $\mathbf{B_0} = B_0 (\sin{\theta_{\rm Bn}} \mathbf{e_y}+\cos{\theta_{\rm Bn}} \mathbf{e_z})$, with inclination angle $\theta_{\rm Bn}$ ranging from $50^\circ$ to $90^\circ$. %Particles are ablated from both sides of a thin solid-density target placed at z = 0. 
%but the simulations were stopped before significant numbers of particles reached the +z boundary. 
The total number of particles per cell is chosen to be 500 at ablation density $n_{\rm e,ab}$. The simulation box is 40000 cells in $z$ and 40 cells in $x$, corresponding to a domain size of $30000 \times 30 d_{\rm e,ab}$, where $d_{\rm e,ab}=c/\omega_{{\rm pe,ab}}$ is the electron skin depth calculated at the ablation density $n_{\rm ab}$. The heating operator is uniform in the transverse direction %The size of the heating operator along x was $L_H = 320 d_{e,ab}$ (i.e. heating is nearly uniform)
 and applied during the first $2\, \Omega_{\rm i}^{-1}$ ($\Omega_{\rm i}\equiv eB_{\rm up}/m_{\rm i}$ is the ion upstream gyrofrequency). The simulations were carried out with a reduced ion-to-electron mass ratio $m_{\rm i}/m_{\rm e} = 100$ (meaning $d_{\rm i,ab} = 10 d_{\rm e,ab}$) and a reduced speed of light compared to the electron thermal speed, $T_{\rm e,ab} / m_{\rm e}c^2 = 0.04$ \citep{FoxPoP2018}. A single ion species plasma with $Z=1$ is considered. The runs cover a range of upstream plasma beta $\beta_{\rm e,up} = 2 n_{\rm e,up} T_{\rm e,up} / B_{\rm up}^2 = 0.5-2$ through varying the upstream temperature with fixed upstream density $n_{\rm up}/n_{\rm ab}=0.05$. We conduct both collisionless and modestly collisional runs, where the collisionality is parameterized by { $\Lambda_{\rm up} \equiv\lambda_{\rm e,mfp}^{\rm th}/d_{\rm i,up} \sim 0.01-0.34$, where $\lambda_{\rm e,mfp}^{\rm th}$ is the mean free path calculated for an electron traveling in upstream plasma with $T_{\rm up}$ and $n_{\rm up}$}. The shock speed is typically $M_{\rm A}=v_{\rm sh}/V_{\rm A,up} \sim 15$ and $M_{\rm e}^{\rm th}=v_{\rm sh}/v_{\rm e,up}^{\rm th} \sim 2.8$, where $v_{\rm e,up}^{\rm th}=\sqrt{2 T_{\rm e,up}/m_{\rm e}}$ is the upstream electron thermal speed. %In addition, we have conducted a set of two-slab simulations, which are easier to interpret and compare to other numerical works (e.g. \cite{twoslab}) on magnetized collisionless shocks, with $M_A \sim 7 - 31$, $\beta_{\rm e,up} \sim 1$, $M_{Si} \sim 2-8$, also covering 1D/2D and $m_i/m_e =100,\, 400,\, 900$. Box size for these simulations is typically $5120 \times 20 d_{\rm e,up}$, having $25600 \times 100$ nodes and 20-100 particles per cell. For both ablation and two-slab, 
 We conduct simulations for $\sim 8 \Omega_{\rm i}^{-1}$, which is sufficient to observe shock formation and the initial stages of particle acceleration. We also tag particles that originated from the ablating foil (``piston'' particles) and from the ambient magnetized plasma (``background'' particles) in order to clarify the physics of piston-driven collisionless shocks.

\begin{figure}
    \centering
    \includegraphics[width=8cm]{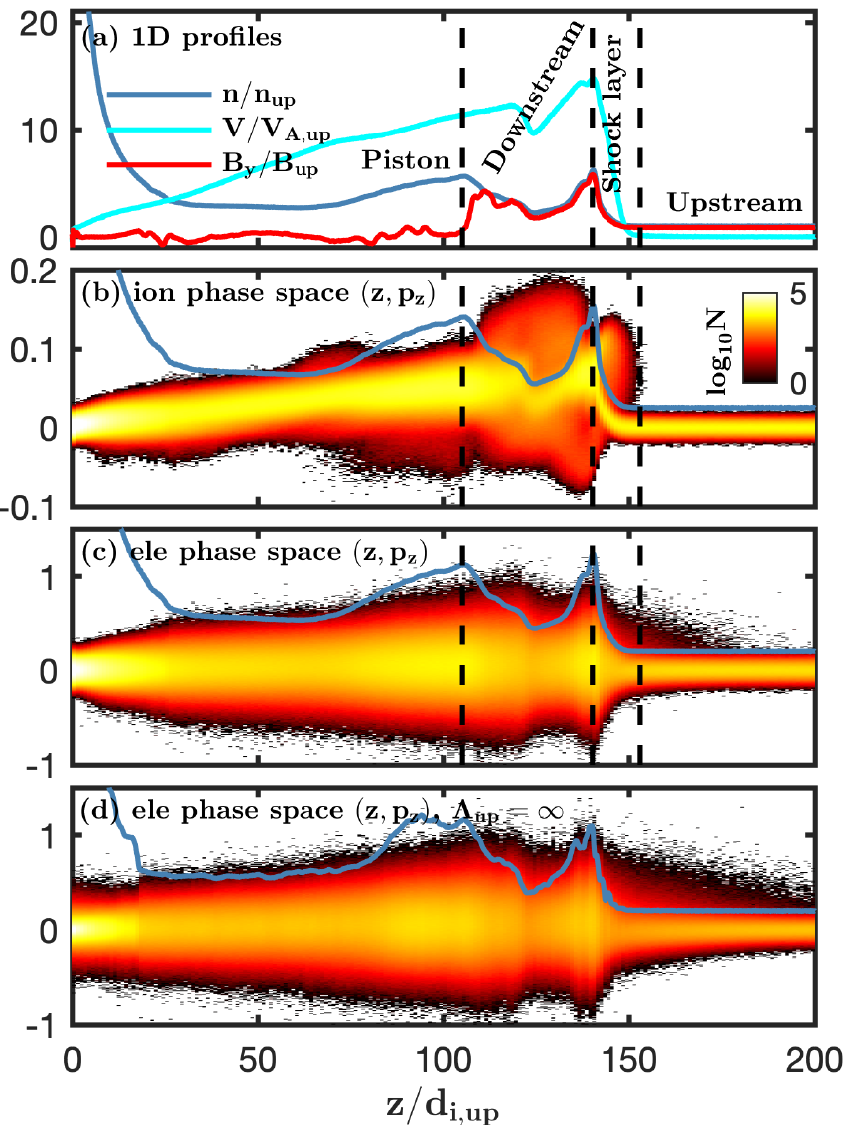}
    \caption{Structure of an ablation driven shock with accelerated electron population at $M_{\rm A}\sim15$, $\beta_{\rm e,up}\sim1$, $\theta_{\rm Bn}=60^\circ$, $m_{\rm i}/m_{\rm e}=100$, $\Lambda_{\rm up}=0.34$ shock at $\Omega_{\rm i}t=8$. Transversly averaged 1D profiles of plasma density, $B_{\rm y}$, and flow speed (a), ion (b) and electron (c) $z-p_{\rm z}$ phase distributions and ion density profiles. (d) electron $z-p_{\rm z}$ phase distribution for a similar collisionless run. Dashed vertical lines separate shock regions - piston, downstream, shock layer, and upstream.}
\end{figure}

%\newpage
%\section{Ablation 1D/quasi-1D simulation shock structure}

%Below, we will discuss the structure of the magnetized collisionless shock in ablation simulation with $M_A \sim 15$, $\beta_{\rm e,up} = 1$, $M_{Si} \sim 2$, $\lambda_{e,mfp}= 6.7 d_{\rm i,up}$. 
Figure 1a shows transversly averaged 1D profiles at $\Omega_{\rm i}t=8$ for simulation with $M_{\rm A} \sim 15$, $\beta_{\rm e,up} = 2$, $M_{\rm e}^{\rm th} \sim 2.8$, and $\Lambda_{\rm up}= 0.34$, which exhibits electron pre-acceleration. This shock is self-consistently formed by a piston plasma expanding into the ambient magnetized plasma and requires a few ion gyrotimes to separate from the piston. Here, we define the shock regions as follows. The piston is defined as the region from the target (z=0) to the edge of the magnetic cavity ($z/d_{\rm i,up}=105$); the shock layer is defined as the region between the overshoot peak ($z/d_{\rm i,up}=140$) and the location where ion gyration stops {($z_{\rm up}/d_{\rm i,up} \equiv (z_{\rm shock}+\rho_{i})/d_{\rm i,up}=153$)}; the downstream and upstream appear to the left and to the right from the shock layer. Here, $\rho_i \approx 13 d_{\rm i,up}$ is the ion gyroradius at the shock front. The jump ratios for magnetic field, density, and electron and ion temperatures are $B_{\rm down}/B_{\rm up} \approx 4$, $n_{\rm i,down}/n_{\rm i,up} \approx 4$, $T_{\rm e,down}/T_{\rm e,up}\approx 20$, and $T_{\rm i,down}/T_{\rm i,up}\approx 35$, which is in approximate agreement with the Rankine-Hugoniot MHD jump conditions in the $M_{\rm A} \gg 1$ limit \citep{FITZPATRICK}, and which indicate the formation of a shock. % - $n_{\rm down}/n_{\rm up}\equiv r \approx (\Gamma+1)/(\Gamma-1) \approx 4$, $T_{\rm down}/T_{\rm up}\equiv (1+\Gamma M_{A}^2(1-r^{-1})+\beta_{up}^{-1}(1-r^2))/r \approx 31$ with $\Gamma=5/3$ being an adiabatic gas constant of 3D gas (since $T_{e,\perp}/T_{e,\parallel}\approx 1$ and $T_{i,\perp}/T_{i,\parallel}\approx 2-3$ in the downstream region in simulations with $\theta_{Bn}=60^\circ$ regardless of collisionality). Thus, we indeed observe a fully formed shock structure.
%and reformation process (variations of magnetic field/density overshoot) is synchronized for both magnetic field and density overshoots (unlike \cite{UmedaDaicho2018}, probably, due to restriction to quasi-1D model), with the same reformation period $\approx 1.8\, \Omega_{\rm i,up}^{-1}$, in agreement with other numerical works, e.g., \cite{twoslab}. Overshoot variation amplitude is much smaller than in auxiliary simulations with two-slab collisionless shock model. 
Figure 1b shows the ion $z-p_{\rm z}$ phase space distribution, with the blue line representing the ion density profile. Here we observe ion reflection in the shock layer near $z \approx 145 d_{\rm i,up}$. In this quasi-perpendicular shock with $\theta_{\rm Bn}=60^\circ$, ions are not reflected far upstream, gyrating with $\rho_{\rm i}= M_{\rm A}\, d_{\rm i,shock} \approx \, 13 \, d_{\rm i,up}$. %, being attached to the shock layer, which is fairly consistent with our definition of shock layer, $l_{\rm shock}\approx 9\, d_{\rm i,up}$. 
 %Ion gyration is also seen in the downstream at $z/d_{\rm i,up}=130$. 
% The relative velocity of reflected ions in the shock layer is $\approx 5 C_{\rm S,ab}$ ($C_{\rm S,ab}=\sqrt{T_{e,ab}/m_i}$ is the isothermal sound speed), which is larger than the upstream electron thermal ($2.25 C_{\rm S,ab}$) and Alfv{\'e}n ($\sim0.225 C_{\rm S,ab}$) velocities. This may lead to multi-stream instabilities in the shock layer \cite{Umeda2014}.
%\re{reference? more discussion on waves?}. 
 The relative velocity of reflected ions in the shock layer is $\approx 2.2\, v_{\rm e,up}^{\rm th}$, which is larger than the upstream electron thermal and Alfv{\'e}n ($\sim0.22\, v_{\rm e,up}^{\rm th}$) velocities. This may lead to multi-stream instabilities in the shock layer \citep{Umeda2014}.
%Since the role of collisions for electrons is vanishingly small on $d_i$-scales, these instabilities and corresponding waves are the only possible primary source for the flow-to-heat energy transfer, and may also be responsible for electron pre-energization. 

Electron energization in the shock layer and upstream is also observed in the electron $z-p_{\rm z}$ phase space distribution, Fig. 1c, around $z/d_{\rm i,up} \approx 150 - 160$. These fast particles have significantly enhanced momentum and $\sim 100\times$ energy compared to the thermal upstream particles, and may ultimately start the cyclic DSA process \citep{Rui2019}. {A similar effect was observed in prior 1D/2D PIC simulations \citep{AmanoHoshino2007,AmanoHoshino2009,Guo2014} with shock parameters close to those presented in this work} and was interpreted as a combination of Shock Surfing Acceleration (SSA), in which electrons are pre-accelerated by electrostatic solitary waves formed in the shock foot region by multi-stream instabilities, and Shock Drift Acceleration (SDA), in which electrons are reflected by the shock magnetic overshoot \citep{REVIEW2}. Collisions play an important role in the electron pre-energization, as we see in Figures 1c,d, where the smaller energized fraction is evident in a collisional case (Fig 1c), in contrast to an identical collisionless run (Fig 1d). %as may be seen in Figure 1d, where electron $z-p_z$ phase plot for identical collisionless run is shown. % was interpreted as a combination/solely as of Shock Surfing Acceleration (SSA) energization mechanism (see Chapter 5.4.2 in \cite{REVIEW2}) due to pre-acceleration of electrons by electrostatic solitary waves formed in the foot by multi-stream instabilities and Shock Drift Acceleartion (SDA), which is basically a reflection of electron by magnetic overshoot of the shock. 
%Here, we performed a comprehensive parameteric search in order to carry SDA pre-acceleration mechanism from astrophysical to laboratory expanding plasma conditions.

\begin{figure}
    \centering
    \includegraphics[width=8cm]{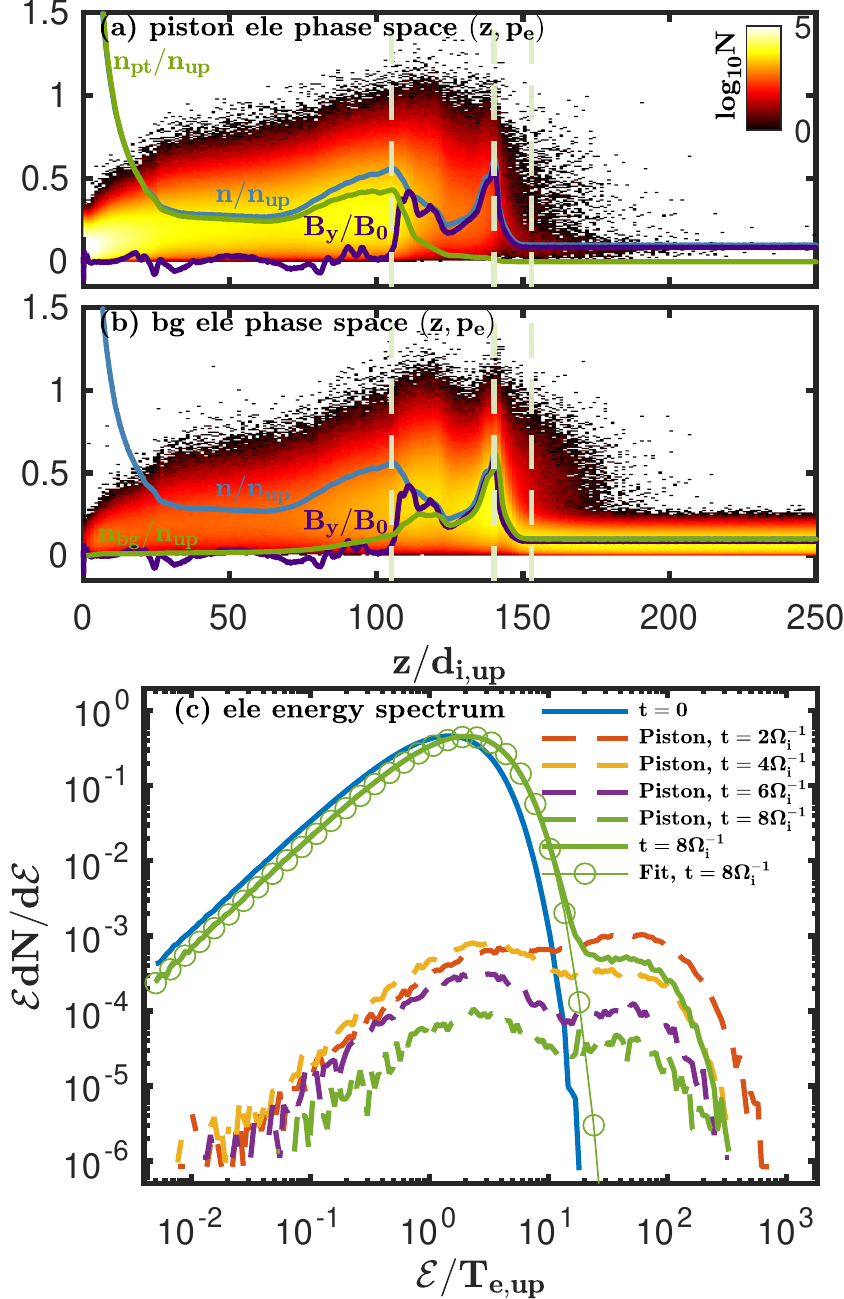}
    \caption{{$z-p_{\rm e,z}$ phase space distribution for particles tagged as (a) piston and (b) background electrons, (c) upstream electron energy spectrum {for the same run as in Figure 1}. Dashed lines in (c) show the piston-tagged particle energy spectrum in the upstream at $t\sim 2,4,6,8\, \Omega_{\rm i}^{-1}$, and the circled line is the fit of bulk part of the upstream spectrum at $\sim 8\, \Omega_{\rm i}^{-1}$.}}
\end{figure}

\begin{figure*}
    \centering
    \includegraphics[width=16cm]{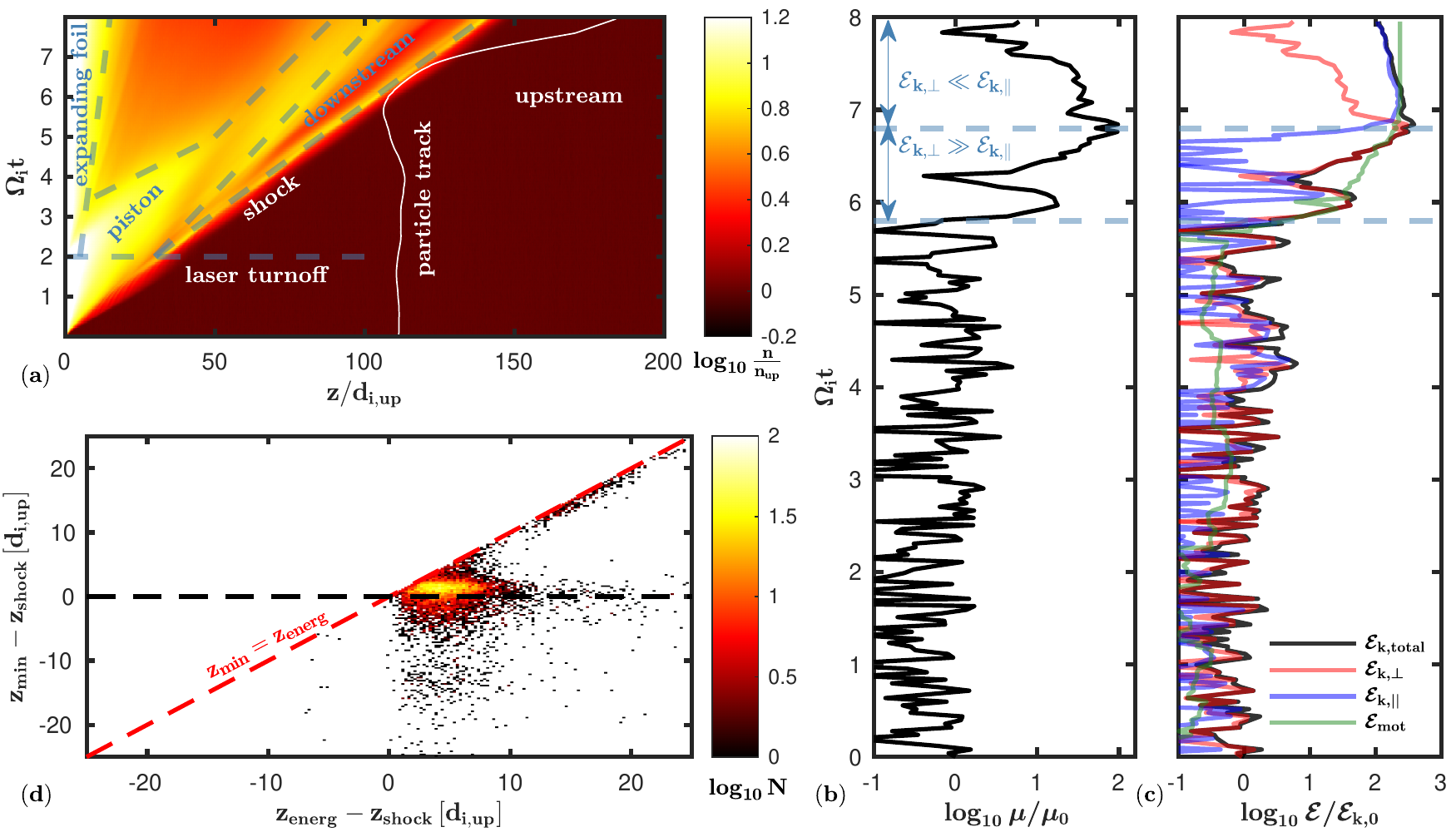}
    \caption{Trajectory of the energized electron (solid white line) in (a) density profile evolution over time, (b) first adiabatic invariant normalized to its initial value, and {(c) total kinetic energy (black solid line), perpendicular kinetic energy (red dotted line), parallel kinetic energy (blue dotted line), energy deposited onto particle by motional electric field (green dashed line)} evolution of particle normalized to initial electron energy. (d) 2D histogram of energized ($p_{\rm e}/m_{\rm e}c>0.3$) particles reflected into the upstream ($z>z_{\rm up}$) at the end of simulation in ($z_{\rm energ}-z_{\rm shock}$, $z_{\rm min}-z_{\rm shock}$) coordinates. The red dashed line demarkates $z_{\rm min} <= z_{\rm energ}$, which is required by definition. %SDA mechanism is acting as particle energy source and reflection mechanism.
    }
\end{figure*}

%\section{Electron energization}

Figure 2 shows in greater detail the time history of electron acceleration. Figures 2a  and  2b show the electron phase space for piston- and background-tagged electrons at $\Omega_{\rm i}t=8$, respectively. The same shock structure regions as above are specified here with dashed lines.
%, with only difference of a $z=0$ point co-located with the centre of an ablating foil.
Blue curves indicate the total electron density profiles, while the green lines indicate the piston electron (Fig 2a) or background electron (Fig 2b) density profiles. %The piston particles, originating from the solid density foil due to ablation process, weakly propagate into the downstream from the left due to collisional effects -- 
Due to collisions, the piston electrons are collisionally slowed in the ambient plasma and are largely stopped before the shock. In contrast, in collisionless simulations, we observe a strong electron bunch which is able to trigger waves in the shock layer (see Figure 1d). Figures 2a and 2b show that the whole shock structure (downstream, shock layer, upstream) is well developed and independent of the piston at this time. %Piston plasma is heated in a non-uniform way, since the heating operator mimicing laser-foil interaction works until $\Omega_{\rm i,up}t=2$ due to finite laser pulse duration. Close to $z=0$, we see a slowly expanding foil which was not heated by laser pulse. In the upstream ($z/d_{\rm i,up}>150$), we see a population of electrons with momentum up to $p_{\rm e,max}/m_ec\approx 0.8$. 

Figure 2c shows the energy spectrum in the upstream ({from $z_{\rm up}$ to the right boundary of the simulation box}) at several times from $\Omega_{\rm i}t = 0$  to  $8$. The contribution to the spectrum from piston-tagged particles is shown with dashed lines. The fit of the bulk part of the late-time electron spectrum is also presented (green solid-circled line). Here, we see that at $t=2\, \Omega_{\rm i}^{-1}$ (the duration of the experiment in \citet{Schaeffer2017}), the non-thermal tail is already there, though the downstream is not yet developed at that time and the non-thermal population is predominantly comprised of piston particles. We find that it requires at least $6\, \Omega_{\rm i}^{-1}$ for the nonthermal tail to be dominated by background particles. %Over time, up until $8\, \Omega_{\rm i,up}^{-1}$, the tail persists with minor collisional deceleration (collisionless simulations do not show such shrinking of the non-thermal electron energy tail). 
The green dashed line, representing the energy spectrum of piston-tagged electrons at $t=8\, \Omega_{\rm i}^{-1}$ in Figure 2c, shows that by this time the piston contribution to the energized particles is small in comparison to the background particles, comprising $<10\%$ for $\mathcal{E}/T_{\rm e,up}$ in the range of 10-100. Electron energization is fairly efficient: the fraction of upstream electrons with energy $\mathcal{E}>20\, T_{\rm e,up}$ is $\epsilon_e \sim 5\%$, in agreement with 1D simulations with similar dimensionless shock parameters \citep{Rui2019}. We convert maximum electron energy to physical units, assuming that it is proportional to the kinetic energy of the shock flow relative to the upstream, $\mathcal{E}_{\rm e,max} \propto m_{\rm i} v_{\rm sh}^2$. Auxiliary simulations with two-slab shock geometry verify this scaling. For $v_{\rm sh} = 700\, \rm km/s$ (typical laboratory speeds), $\mathcal{E}_{\rm e,max} \approx 11\, \rm keV$.

Figure 3 illustrates the mechanism for electron energization for an electron which ends up in the upstream. Figure 3a shows the evolution of the density profile over time superposed with a particle trajectory near the shock. It shows the evolution of the shock structure (dashed line labeled `shock'), expansion of the ablating foil (white area labeled `expanding foil'), propagation of the piston (yellow area labeled `piston'), and development of the contact discontinuity and shock downstream (starting from $\sim3\, \Omega_{\rm i}^{-1}$). The white line shows a particle track in $(z,t)$ space. During the first $5 \, \Omega_{\rm i}^{-1}$, the electron quivers around $z/d_{\rm i,up}\sim110$ with nearly constant magnetic moment $\mu \equiv v_{e\perp}^2/B$ (Figure 3b) and energy (Figure 3c), and once the electron gets within $\sim 1-10\, d_{\rm i,up}$ from the overshoot, the electron experiences a non-adiabatic (Fig. 3b) gain of perpendicular energy over a time $\sim \Omega_{\rm i}^{-1}$. {This type of particle energization is consistent with SDA \citep{Guo2014}, which only requires (a) the presense of the motional electic field {being dominantly responsible} for perpendicular energy gain (Fig. 3c) and (b) $\nabla B$ drift in the shock layer ($\nabla B \times \vec{B} \parallel {\bf e_x} \neq 0$, as seen in Figure 1a).} After traveling with the shock front for $\approx 1 \Omega_{\rm i}^{-1}$ at a location within $\sim6-8\, d_{\rm i, up}$ from the overshoot, {the accelerated electron} is reflected from the magnetic overshoot to the upstream, losing its perpendicular energy (Fig. 3c) and escaping along the magnetic field line. Tracking back all energized particles in the upstream (i.e., particles with $p_{\rm e}/m_{\rm e}c>0.3$ and $z>z_{\rm up}$), we estimate where this particle population was accelerated (i.e., where $p_{\rm e}/m_{\rm e}c>0.3$ for the first time throughout the simulation) with respect to the shock, $z_{\rm energ}-z_{\rm shock}$, and how deep these particles get into the shock over the whole shock evolution, $z_{\rm min}-z_{\rm shock}$, Fig. 3d. This analysis indicates that particles are predominantly energized in the shock foot ($z_{\rm energ}-z_{\rm shock} > 0$), rather than in the downstream, and that only a small number of particles even sample the downstream. Average values of these quantities are: $\langle {z_{\rm energ}-z_{\rm shock}}\rangle \approx 5.6 d_{\rm i,up}$ and $\langle {z_{\rm min}-z_{\rm shock}}\rangle \approx 0.8 d_{\rm i,up}$.  %meaning that energized particles do not travel too far into the shock, typically reaching as far as to the shock layer, where energization occurs. 
The mean energy e-folding time of this population is $1.8\, \Omega_{\rm i}^{-1}$, which is again in good agreement with \citep{Guo2014}. {The typical energy gain in SDA, $\Delta \mathcal{E}_{\rm SDA} / T_{\rm e,up} = M_{\rm A}^{-1} (m_{\rm i}/m_{\rm e}) (m_{\rm e}v_{\rm sh}^2/T_{\rm e,up}) \sin{\theta_{\rm Bn}} (\delta x/d_{\rm i,up}) \sim 565$ ($\Delta \mathcal{E}_{\rm SDA}/m_{\rm e}c^2 \sim 1.1$), is fairly consistent with energy gains observed in our simulations ($\delta x \sim 10 d_{\rm i,up}$ is the transverse distance travelled by electron in shock layer before the reflection)}. SSA \citep{AmanoHoshino2007} and cyclic SDA \citep{Guo2014} were not observed in the run, since the waves are suppressed in collisional simulations: $\delta B/B_{\rm up} < 20 \%$ and $E_{\rm es}/B_{\rm up}<0.1$ in the collisional case, in contrast to $\delta B/B_{\rm up} \sim 100 \%$ and $E_{\rm es}/B_{\rm up}\sim 0.3$ in the collisionless run. Here, $\delta B$ is the magnetic field perturbation magnitude and $E_{es}$ is the electrostatic component of the electric field. %The collisionless run may in principle deliver cyclic SDA (see Figure 1d), where both signs of energized $p_{e,z}$ in the upstream are seen.
%\re{more details about waves? more details about energy gain?}

{Figure 4a-b summarizes the whole set of our collisionless (blue) and collisional (red) ablation simulations with $M_{\rm A} \sim 15$ and $\beta_{\rm e,up}\sim 1$. We varied the shock angle $\theta_{\rm Bn}$, collisionality ($\Lambda_{\rm up}$), and Alfv{\'e}n Mach number $M_{\rm A}$, and observe significant energization of the upstream electron population. We quantify the accelerated electrons in terms of two parameters: $\overline{\mathcal{E}_{\rm e,up}}\equiv \int_{z_{\rm up}}\mathcal{E} f(\mathcal{E}) d\mathcal{E}/\int_{z_{\rm up}}f(\mathcal{E}) d\mathcal{E}$, which is the energy moment of the distribution function $f(\mathcal{E})$ calculated in the upstream; and shock reflectivity $R$, which is the fraction of nonthermal particles in the upstream $R \equiv n_{\rm e,up}(\mathcal{E}>20 \,{T_{\rm e,up}})/n_{\rm e,up}$. Error bars are obtained by varying the analysis window within $5 d_{\rm i,up}$. The green asterisk corresponds to the reference simulation described above. A parametric scan shows a range around $R \approx 1\%$--$2\%$ of nonthermal particles and $\overline{\mathcal{E}_{\rm e,up}}/T_{\rm e,up}\sim 4 - 5$ for $\theta_{\rm Bn}=60^\circ$ for collisional runs. The trend toward smaller number of particles for larger shock angles is in qualitative agreement with similar simulations in \citet{AmanoHoshino2007}. This is tied to the size of the loss cone allowing particles to escape along the magnetic field line when the condition $u_\perp \geq C_{\rm s,up}({B_{\rm up}/B_{\rm overshoot}})^{1/2}$ is satisfied \citep{AmanoHoshino2007}. Here, $u_\perp$ is the perpendicular velocity with respect to local magnetic field and $C_{\rm s,up}$ is the upstream sound speed. The fraction of non-thermal particles saturates for angles smaller than $65^\circ$, which is again in agreement with the analytical prediction from \citet{AmanoHoshino2007}. The fraction of reflected particles is suppressed for collisional runs in comparison to collisionless, but still stays within $R \approx 1\%-2 \%$.  
Figures 4c-f demonstrate a scan on collisionality (Figure 4c,e) and $M_{\rm A}$ (Figure 4d,f). 
They show the robustness of the proposed pre-acceleration mechanism to variations in shock speed for Alfv{\'e}n Mach numbers larger than threshold for injection $M_{A}^{\rm inj}$, $M_{\rm A} \geq M_{\rm A}^{\rm inj}\equiv0.5 \cos{(\theta_{\rm Bn})}(\beta_{\rm e,up}\,m_{\rm i}/m_{\rm e})^{1/2} \approx 3.5$ \citep{AmanoHoshino2010}.} %\re{add simulations with variations in $\beta_e$?}. 
%The same parameters are also shown for simulations with ablation setup - collisionless simulations (blue points and lines) %where the heating operator works throughout the whole simulation 
%and collisional simulations (red markers) with various degree of collisionality ($\lambda_{e,mfp}/d_{\rm i,up}=0.67,\,2.0,\,6.7$). 
{Figure 5a-c presents the collisionality scan for $\Lambda_{\rm up}$ from 0.11 to 0.011, and it clearly shows how the gradual transition to more collisional plasma suppresses the population of energetic electrons in the upstream. The collisionality threshold criteria is found to be $\Lambda_{\rm up}\gg0.01$.} %To explain this, we consider the Dreicer electric field ($E_{\rm D} \equiv T_{\rm e,up}/\lambda_{\rm e,mfp}^{\rm th}$), which is a condition for runaway of accelerated particles. If we require the shock electric field $E \sim v_{\rm sh} B$ to be comparable to or exceed the Dreicer field, then we may rewrite this condition in terms of $M_{\rm A}$ and $\beta_{\rm e,up}$, $\Lambda_{\rm up} >\beta_{\rm e,up}/M_{\rm A}$, or, for our values of $M_{\rm A}$ and $\beta_{\rm e,up}$, $\Lambda_{\rm up}>0.06$, which is close to the numerically obtained criterion. 
%Green lines on Fig 5a-c illustrate the transition from below to above the Dreicer electric field value in the shock layer.} 
%Thus, we have demonstrated that astrophysically-relevant mechanisms of electron pre-acceleration in magnetized collisionless shocks are feasible to observe in expanding laboratory laser plasmas, and the pre-acceleration mechanism is stable against moderate variations in Mach numbers and collisionality.

\begin{figure}
    \includegraphics[width=8cm]{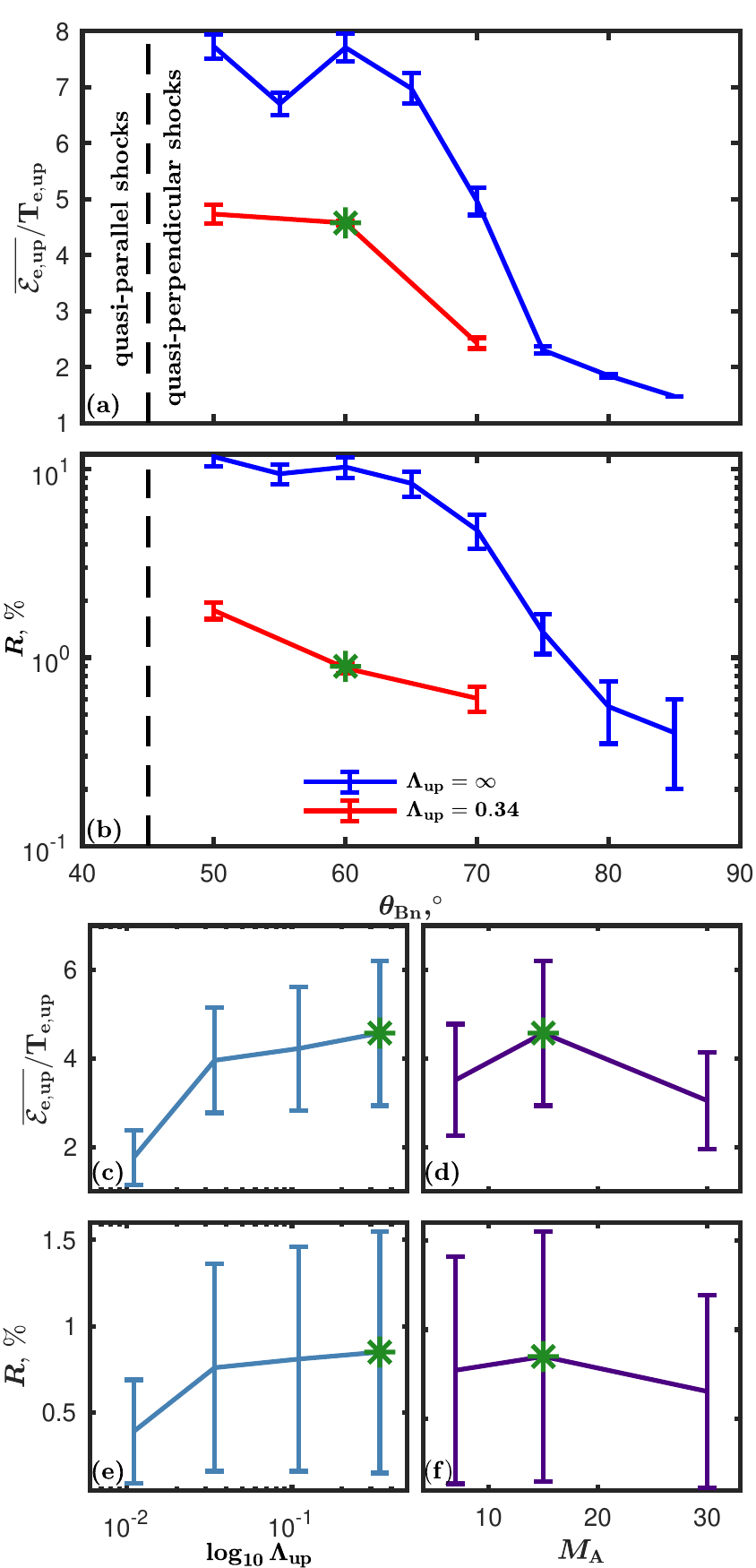}
    \caption{Dependence of the properties of the nonthermal electron population - (a) $\overline{\mathcal{E}_{\rm e,up}}/T_{\rm e,up}$ and (b) $R$ -- on shock angle $\theta_{\rm Bn}$ for collisionless (blue) and collisional (red) simulations; scans on (c,e) $\Lambda_{\rm up}$  and (d,f) $M_{\rm A}$ for collisional simulations. Green asterisk corresponds to the reference simulation described in Fig. 1-3.}  %collisionless two-slab simulation (green and red triangle). Dependence of $N_{\rm nth}/N_{\rm up}, T_{\rm e,nth}/T_{\rm e,up}$ on $\lambda_{e,mfp}/d_{\rm i,up}$ (c,e) and (d,f) $M_A$.}
\end{figure}

\begin{figure}
    \centering
    \includegraphics[width = 8cm]{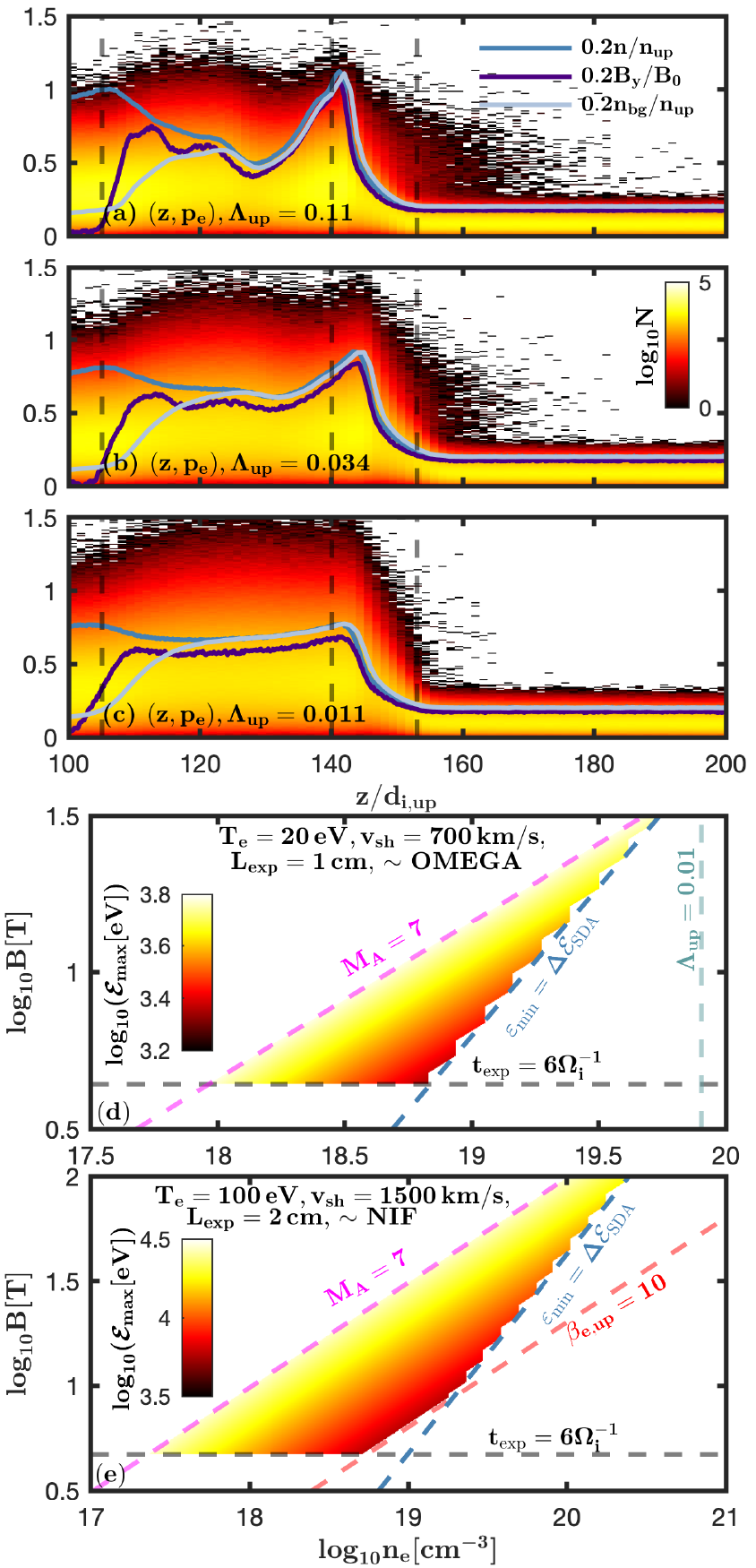}
    \caption{{Electron phase space distribution in ($z/{d_{\rm i,up}}, \, p_e/m_{\rm e}c$) coordinates for the parameters of the run presented above with the collisionality level $\Lambda_{\rm up}$ being (a) 0.11, (b) 0.034, (c) 0.011. The gradual drop of the maximum energies and pre-accelerated electron numbers is seen for larger collisionalities. Colormap of the maximum electron energy achievable by collisionless shock via SDA that may be detected in the experimental setup simular to (d) OMEGA and (e) NIF.}}
\end{figure}

{Let us summarize the experimental requirements that will allow us to study shock acceleration of electrons. (1) \textit{Shock parameters} -  we need the shock parameters to be in the right regime for the efficient SDA manifestation, which, according to our simulations and data from literature, requires $33 \geq M_{\rm A} \geq 7 \geq M_{\rm inj}$ and $\beta_{\rm e,up}\sim 1$. The shock angle is also important, as only quasi-perpendicular shocks with $\theta_{\rm Bn}<70^\circ$ show significant acceleration. (2) \textit{System size} - we require the spatial size of the setup to be large enough, so the shock will have enough time to develop and accelerate electrons. This is a constraint on both the experiment time $t_{\rm exp}$ and system size $L_{\rm exp}$, which are related by $t_{\rm exp}=L_{\rm exp}/ v_{\rm sh}$, where $v_{\rm sh}$ is the shock speed. We find that the shock develops in $\sim 2\Omega_i^{-1}$, that the timescale for particle acceleration is $\sim 1.8 \Omega_i^{-1}$. However, we then find $\sim 6\Omega_i^{-1}$ are required for the background-accelerated particles to dominate, allowing a clean detection of acceleration. %Here, $t_{\rm exp}=L_{\rm system}/v_{sh}$ is the maximum experimental time, $L_{\rm system}$ is the spatial scale of the system, and $v_{sh}$ is the shock speed. 
(3) \textit{Collisionality} - another condition for the experimental observation of non-thermal electrons is sufficiently low collisionality. This requirement can be naturally separated into two subcategories: (I) sufficiently low collisionality for the SDA to effectively accelerate electrons (%necessary condition is that the collision rate is smaller than acceleration rate $\nu_{ee}/\Omega_i \leq 1$; 
our numerical analysis implies that the sufficient condition is $\Lambda_{\rm up}\gg0.01$) and (II) sufficiently low collisionality for the accelerated electrons to leave the shock layer and reach the detector without significant energy losses, i.e. the mean free path of energetic electron is larger than the system size ($\lambda_{\rm e,mfp}^{\varepsilon}/L_{\rm exp}\geq 1$). Another condition that is related to collisonality is that $\mathcal{E}_{\rm max}$, the maximum electron energy that can be achieved by the given collisionless shock via SDA \citep{Guo2014}, is no less than the minimum electron energy that satisfies escape conditions, $\varepsilon_{\rm min}$: $\mathcal{E}_{\rm max} \geq \varepsilon_{\rm min}$. 

Figures 5d,e illustrate the parameter space $(n,B,\mathcal{E}_{\rm max})$ which satisfies the conditions listed above for typical parameters of experimental setups at large laser facilities like OMEGA and NIF. For typical experimental parameters at the OMEGA facility (system size $L_{\rm exp} = 1\, {\rm  cm}$, background plasma temperature $T_{\rm e}\sim 20\, {\rm eV}$,  shock speed $v_{\rm sh}=700\, \rm km/s$ \citep{Schaeffer2017}), {observation of non-thermal electrons} requires a regime with $B \sim 10\, \rm T$ magnetic field and upstream plasma density $n_e \sim 10^{18}-10^{19}\, {\rm cm^{-3}}$. In this case, we expect the electrons of energies between $\varepsilon_{\rm min}=1.5 \, \rm keV$ and $\mathcal{E}_{\rm max} = 5 \, \rm keV$ to escape the experimental setup and be available for observation. These parameters are already available at the OMEGA facility (e.g., magnetic fields of around 15 T were previously reported in \cite{Fiksel2015}). For NIF-like parameters ($L_{\rm exp}=2\, \rm cm$, $T_{\rm e} =100 \, \rm eV$, $v_{\rm sh}=1500\, \rm km/s$), a regime with $B \sim 10^{1}\, \rm T$ and $n_e \sim 10^{18}-10^{19}\, {\rm cm^{-3}}$ is needed, allowing observations of particles in the range from 1.5 to 10 keV; our PIC simulations demonstrate that such high electron energies are achievable. While magnetized collisionless shock experiments have not yet been conducted at NIF, these values are reasonable extensions from OMEGA experiments to a larger laser facility like NIF. Some parameters, such as temperature and flow speed, were recently reported for the experimental study of a Weibel shock at NIF \citep{Fiuza2020}.}

{It is useful to note that we do not expect a significant influence of the shock curvature on the SDA acceleration of electrons. In principle, the shock front curvature does affect the efficiency of the SDA, since it requires a significant transverse motion of the pre-accelerated particle. But in our case, at later stages of shock evolution, the radius of curvature scales as $M_{\rm A} \Omega_i t$ and, after {a few ion gyrotimes}, will be significantly larger than transverse acceleration scales within the shock. Thus, the SDA mechanism will not be affected.}% We performed simulations that vary the heating radius with respect to the transverse size of the box and they confirm this finding.}

It is also important to note that we conduct 2D simulations with `out-of-plane' magnetic field (i.e., with significant magnetic field component along $\bf e_y$, which is perpendicular to the simulation box plane $\bf \{ e_x, e_z\}$), which is known to affect the structure of the shock itself \citep{twoslab1,twoslab2}, as well as the electron energization efficiency \citep{Guo2014,Crumley2019}. Our auxiliary simulations suggest that out-of-plane runs demonstrate at least an order of magnitude advantage in shock reflectivity $R$ in comparison to in-plane runs. Resolving this will require 3D simulations, which are too computationally demanding at present, and, of course, experimental measurements. {In order to study the convergence of our results, we conducted a few auxilliary runs with $M_{\rm A}\sim15,\,\beta_{\rm e,up}\sim1,\,\Lambda_{\rm up}\sim 0.34$, and $\theta_{\rm Bn}=60^\circ$, varying transverse size (strictly 1D, 2$\times$ and 4$\times$ the transverse size of the box mentioned above), heating radius (so the heating radius equals the transverse size of the box), mass ratio (running a strictly 1D run with $m_{\rm i}/m_{\rm e}=400$), and absolute value of the shock speed relative to the speed of light, and concluded that changing these parameters does not significantly affect $R$ and $\overline{\mathcal{E}_{\rm e,up}}$.}

% (in terms of $n_{\rm e,up}(E>20 \, T_{\rm e,up})/n_{e, up}$).  %Recent 3D simulations \re{ref?} show that the shock reflectivity in 'out-of-plane' magnetic field configuration is closer to the reflectivity in 3D runs than in case of 'in-plane' configuration.

%Short discussion of structure of quasi-1D or 1D shock in ablation case, contrast with slab, RH condition discussion, 1D vs 2D, MMi effects, reformation(?) - Figure 1
%\section{Electron energization in collisionless two-slab case}
%Focus on collisional ablation simulation with finite time of heating operator which shows the mechanism from Amano\&Hoshino ApJ, 2007, working, discussion of spectrum and its fitting with two weighted maxwellian spectrums, scaling to physical units - Figure 2
%\section{Electron energization mechanism in collisionless ablation case}
%Tracking of multiple(?) particle energization in 1D ablation collisionlal simulation, maybe show a plot of electrostatic wave amplitude instread of density-time color map(?) - Figure 3

%\section{Electron energization in realistic simulations - electron phase plot and spectrums in different regions}
%Discussion of how the shock will behave for finite time application of heating operator, shock angle dependence, role of collisions, slab vs ablation, MMi, 1D vs 2D - Figures 4\&5

%\section{Discussion}

In summary, we have conducted a multi-parameter investigation of electron pre-acceleration by collisionless magnetized shocks in experimental conditions of expanding laboratory laser plasmas. Our quasi-1D PIC simulations show that it is possible to generate a population of non-thermal electrons in the upstream and shock layer with energies up to tens of keV when the shock parameters are close to those that were obtained experimentally in
\cite{Schaeffer2017,Schaeffer2019}. We also formulate the experimental conditions needed for the robust observation of electron injection by magnetized collisionless shocks. %This regime of shock parameters $(M_A, \beta_e)$ was obtained in recent experiments, however we find moderately larger systems and/or stronger fields will be needed to robustly observe shock particle acceleration, namely $\sim 6-8 \Omega_i^{-1}$. Our work illustrates experimental requirements for robust observation of shock-energized particles, which may be implemented in the nearest future at OMEGA and NIF facilities. %First, our simulations can run longer than in \citet{Schaeffer2017} ($<2\Omega_{\rm i,up}^{-1}$ vs $8 \Omega_{\rm i,up}^{-1}$) due to the spatial limitation of the experimental setup. Our scan verified that the population of reflected pre-accelerated electrons is a robust effect persisting through variations in both Mach numbers and collisionality. Thus, we may expect that with increased fields, the experiments in \citet{Schaeffer2017} would evolve faster and thus fit within current space constraints. Another important temporal-spatial constraint is a separation of the whole shock structure from the 
%piston (requiring at least $6 \Omega_{\rm i,up}^{-1}$). %as well as escaping of pre-accelerated electrons which will stream along magnetic field far enough so they won't be mixed with the energized electrons of another origin, e.g., electrons accelerated by laser-plasma instabilities. 
%Second, our work argues about the importance of the magnetic field geometry for electron pre-acceleration, giving preference to quasi-perpendicular shocks with $\theta_{Bn}\lesssim 70^\circ$ (in terms of shock reflectivity $R$). {Finally, the collisionality of plasma should be small enough ($\Lambda_{\rm up}\geq 10^{-2}$) so the particles will be able to gain energy from the thermal level and to leave the plasma region without significant energy losses.} 
In the near future, we believe controlled laboratory experiments on electron energization by magnetized collisionless shocks will allow for better understanding of electron energization by moderate-level Alfv{\'e}n Mach number shocks observed in the Earth's magnetosphere, as well as to address the injection problem for high-Mach-number shocks.

Simulations were conducted on the Titan and Summit supercomputers at the Oak Ridge Leadership Computing Facility at the Oak Ridge National Laboratory, supported by the Office of Science of the DOE under Contract No. DE-AC05-00OR22725. This research was also supported by the DOE under Contracts No. DE-SC0014405, DE-SC0016249, DE-NA0003612, and NSF grants PHY-1748958, AST-1814708 and PHY-1804048.


\begin{thebibliography}{}

\bibitem[Amano \& Hoshino(2007)]{AmanoHoshino2007}
Amano T., Hoshino M., 2007, ApJ, 661, 190

\bibitem[Amano \& Hoshino(2009)]{AmanoHoshino2009}
Amano T., Hoshino M., 2009, ApJ, 690, 244

\bibitem[Amano \& Hoshino(2010)]{AmanoHoshino2010}
Amano T., Hoshino M., 2010, Phys. Rev. Lett., 104, 181102

\bibitem[Bohdan et al.(2017)]{twoslab2}
Bohdan A., Niemiec J., Kobzar O., Pohl M., 2017, ApJ, 847, 71

\bibitem[Bell(1978a,b)]{DSA}
Bell A.R., 1978, MNRAS, 182, 147

\bibitem[Bell(1978)]{DSA2}
Bell A.R., 1978, MNRAS, 182, 443

\bibitem[Bulanov(2016)]{Bulanov2016}
Bulanov S.V., 2016, Plasm. Phys. Cont. Fus., 59, 014029

\bibitem[Burgess \& Scholer(2015)]{REVIEW3}
Burgess D., Scholer M., 2015, {\it Collisionless Shocks in Space Plasmas}, Cambridge University Press

\bibitem[Crumley et al.(2019)]{Crumley2019}
Crumley P., Caprioli D., Markoff S., Spitkovsky A., 2019, MNRAS, 485, 5105

\bibitem[Dong et al.(2012)]{Shenguang1}
Dong Q.L., Wang S.J., Lu Q.M., Huang C., Yuan D.W., Liu X., Liu X.X., Li Y.T., Wei H.G., Zhong J.Y., Shi J.R., 2012, Phys. Rev. Lett. {108}, 215001

\bibitem[Fiksel et al.(2015)]{Fiksel2015}
Fiksel G., Agliata A., Barnak D., Brent G., Chang P.Y., Folnsbee L., Gates G., Hasset D., Lonobile D., Magoon J., Mastrosimone D., 2015, Rev. Sc. Inst. { 86}, 016105

\bibitem[Fiuza et al.(2020)]{Fiuza2020}
Fiuza F., Swadling G.F., Grassi A., Rinderknecht H.G., Higginson D.P., Ryutov D.D., Bruulsema C., Drake R.P., Funk S., Glenzer S., Gregori G., C. K. Li, B. B. Pollock, B. A. Remington, J. S. Ross, W. Rozmus, Y. Sakawa, A. Spitkovsky, S. Wilks \& H.-S. Park, 2020, Nat. Phys. 16, 916

\bibitem[Fitzpatrick(2014)]{FITZPATRICK}
Fitzpatrick R., 2014, {\it Plasma Physics. An Introduction.}, CRC Press

\bibitem[Fox et al.(2013)]{Fox2013}
Fox W., Fiksel G., Bhattacharjee A., Chang P.Y., Germaschewski K., Hu S.X. and Nilson P.M., 2013, Phys. Rev. Lett. { 111}, 225002

\bibitem[Fox et al.(2018)]{FoxPoP2018}
Fox W., Matteucci J., Moissard C., Schaeffer D.B., Bhattacharjee A., Germaschewski K., Hu S.X., 2018, Phys. Plasmas { 25}, 102106


\bibitem[Germaschewski et al.(2016)]{Germaschewski2016}
Germaschewski K., Fox W., Abbott S., Ahmadi N., Maynard K., Wang L., Ruhl H., Bhattacharjee A., 2016, J. Comp. Phys. { 318}, 305


\bibitem[Glasmacher et al.(1999)]{COSMICRAYS}
Glasmacher M.A.K., Catanese M.A., Chantell M.C., Covault C.E., Cronin J.W., Fick B.E., Fortson L.F., Fowler J.W., Green K.D., Kieda D.B., Matthews J., 1999, Astroparticle Phys. 10, 2911302

\bibitem[Guo, Sironi \& Narayan(2014)]{Guo2014}
Guo X., Sironi L., Narayan R., 2014, ApJ, { 794}, 153

\bibitem[Huntington et al.(2015)]{Huntington2015}
Huntington C.M., Fiuza F., Ross J.S., Zylstra A.B., Drake R.P., Froula D.H., Gregori G., Kugland N.L., Kuranz C.C., Levy M.C., Li C.K., 2015, Nature Phys., { 11}, 173

\bibitem[Katou \& Amano(2019)]{Katou2019}
Katou T., Amano T., 2019, ApJ, { 874}, 119

\bibitem[Krymsky(1977)]{Krymsky1977}
Krymsky G.F., ASR USSR, 1977, 234, 1306

\bibitem[Masters et al.(2013)]{MastersNPHYS2013}
Masters A., Stawarz L., Fujimoto M., Schwartz S.J., Sergis N., Thomsen M.F., Retino A., Hasegawa H., Zieger B., Lewis G.R., Coates A.J., 2013, Nature Phys. { 9}, 164
\bibitem[Matsumoto, Amano \& Hoshino(2012)]{Matsumoto2012}
Matsumoto Y., Amano T., Hoshino M., 2012, ApJ, { 755}, 109
\bibitem[Matsumoto, Amano \& Kato(2015)]{Matsumoto2015}
Matsumoto Y., Amano T., Kato T. N., Hoshino M., 2015, Science { 347}, 974
\bibitem[Nilson et al.(2006)]{VULCAN}
Nilson P.M., Willingale L., Kaluza M.C., Kamperidis C., Minardi S., Wei M.S., Fernandes P., Notley M., Bandyopadhyay S., Sherlock M., Kingham R.J., 2006, Phys. Rev. Lett. { 97}, 255001

\bibitem[Rosenberg et al.(2015a,b)]{Rosenberg}
Rosenberg M.J., Li C.K., Fox W., Zylstra A.B., Stoeckl C., S{\'e}guin F.H., Frenje J.A., Petrasso R.D., 2015, 
Phys. Rev. Lett. { 114}, 205004
\bibitem[Rosenberg et al.(2015a,b)]{Rosenberg2}
Rosenberg M.J., Li C.K., Fox W., Igumenshchev I., S{\'e}guin F.H., Town R.P.J., Frenje J.A., Stoeckl C., Glebov V., Petrasso R.D., 2015, Nat. Commun. { 6}, 6190

\bibitem[Schaeffer et al.(2017a,b)]{Schaeffer2017}
Schaeffer D.B., Fox W., Haberberger D., Fiksel G., Bhattacharjee A., Barnak D.H., Hu S.X., Germaschewski K., 2017, Phys. Rev. Lett. { 119}, 025001
\bibitem[Schaeffer et al.(2017a,b)]{Schaeffer2017b}
Schaeffer D.B., Fox W., Haberberger D., Fiksel G., Bhattacharjee A., Barnak D.H., Hu S.X., Germaschewski K., Follett R.K., 2017, Phys. Plasmas { 24}, 122702

\bibitem[Schaeffer et al.(2019)]{Schaeffer2019}
Schaeffer D.B., Fox W., Follett R.K., Fiksel G., Li C.K., Matteucci J., Bhattacharjee A., Germaschewski K., 2019, Phys. Rev. Lett. { 122}, 245001

\bibitem[Schaeffer et al.(2020)]{SchaefferPoP2020}
Schaeffer D. B., Fox W., Matteucci J., Lezhnin K.V., Bhattacharjee A, Germaschewski K., 2020, Phys. Plasmas { 27}, 042901

\bibitem[Treumann(2009)]{REVIEW2}
Treumann R. A., 2009, A.\&A. Rev. { 17}, 409

\bibitem[Umeda et al.(2014)]{Umeda2014}
Umeda T., Kidani Y., Matsukiyo S., Yamazaki R., 2014, Phys. Plasmas { 21}, 022102

\bibitem[Umeda et al.(2019)]{UMEDA2019}
Umeda T., Yamazaki R., Ohira Y., Ishizaka N., Kakuchi S., Kuramitsu Y., Matsukiyo S., Miyata I., Morita T., Sakawa Y., Sano T., 2019, Phys. Plasmas { 26}, 032303

\bibitem[Wieland et al.(2016)]{twoslab1}
Wieland V., Pohl M., Niemiec J., Rafighi I., Nishikawa K.-I., 2016, ApJ, { 820}, 62

\bibitem[Xu et al.(2020)]{Rui2019}
Xu R., Spitkovsky A., Caprioli D., 2020, ApJL { 897}, L41

\bibitem[Yamada, Kulsrud \& Ji(2010)]{REVIEW1}
Yamada M., Kulsrud R., Ji H., 2010, Rev. Mod. Phys. { 82}, 603

\bibitem[Zhong et al.(2010)]{Shenguang2}
Zhong J., Li Y., Wang X., Wang J., Dong Q., Xiao C., Wang S., Liu X., Zhang L., An L., Wang F., 2010, Nat. Phys. { 6}, 984

\end{thebibliography}
\end{document}